# A novel hotspot of gelsolin instability and aggregation propensity triggers a new mechanism of amyloidosis

Michela Bollati[1], Luisa Diomede[2], Toni Giorgino[1], Carmina Natale[2], Elisa Fagnani[1], Irene Boniardi[1], Alberto Barbiroli[3], Rebecca Alemani[1], Marten Beeg[2], Marco Gobbi[2], Ana Fakin[4], Eloise Mastrangelo[1], Mario Milani[1], Gianluca Presciuttini[5], Edi Gabellieri[5], Patrizia Cioni[5], Matteo de Rosa[1*]

[1]Istituto di Biofisica, Consiglio Nazionale delle Ricerche, Milano, Italy

[2]Department of Molecular Biochemistry and Pharmacology, Istituto di Ricerche Farmacologiche Mario Negri IRCCS, Milano, Italy

[3]Dipartimento di Scienze per gli Alimenti, la Nutrizione e l'Ambiente, Università degli Studi di Milano, Milano, Italy

[4]Eye Hospital, University Medical Centre Ljubljana, Ljubljana, Slovenia

[5]Istituto di Biofisica, Consiglio Nazionale delle Ricerche, Pisa, Italy

*to whom correspondence should be addressed: matteo.derosa@ibf.cnr.it

**Author Contributions:** M.B, L.D., T.G., and M.d.R conceived and designed the experiments; M.B., L.D., T.G., C.N., E.F., I.B., A.B., M.Be., R.A., G.P., E.G., and M.d.R. collected the data; M.B., L.D., T.G., M.Be., M.M., E.G, P.C., and M.d.R analysed and interpreted the data; M.B., L.D., T.G., and M.d.R. drafted the article; all authors critically revised the article.

**Competing Interest Statement:** There are no conflicts of interest to disclose.

**ABSTRACT (231 of 250 words)**

The multidomain protein gelsolin (*GSN*) is composed of six homologous modules, sequentially named G1 to G6. Single point substitutions in this protein are responsible for AGel amyloidosis, a hereditary disease characterized by progressive corneal lattice dystrophy, cutis laxa, and polyneuropathy. Several different amyloidogenic variants of *GSN* have been identified over the years, but only the most common D187N/Y mutants, in G2, have been thoroughly characterized, and the underlying functional mechanistic link between mutation, altered protein structure, susceptibility to aberrant furin cleavage and aggregative potential resolved. Little is known about the recently identified mutations A551P, E553K and M517R hosted at the interface between G4 and G5, whose aggregation process likely follows an alternative pathway. We demonstrate that these three substitutions impair temperature and pressure stability of GSN but do not increase its susceptibility to furin cleavage, the first event of the canonical aggregation pathway. The variants are also characterized by a higher tendency to aggregate in the unproteolysed forms and show a higher proteotoxicity in a *C. elegans*-based assay. Structural studies point to a destabilization of the interface between G4 and G5 due to three different structural determinants: β-strand breaking, steric hindrance and/or charge repulsion, all implying the impairment of interdomain contacts. All available evidence suggests that the rearrangement of the protein global architecture triggers a furin-independent aggregation of the protein, supporting the existence of a non-canonical pathway of gelsolin amyloidosis pathogenesis.

**SIGNIFICANCE STATEMENT (122 of 120 words)**

Gelsolin amyloidosis is a rare disease typically caused by pathogenic variants in the gelsolin gene. The molecular events leading to the aggregation of gelsolin and formation of pathological amyloids in organs and tissues are still unknown for the numerous pathogenic variants recently discovered. Furthermore, no specific pharmacological therapy for the disease is available. We present a molecular characterization of three novel amyloidogenic substitutions that cluster at the interface between the fourth and fifth domains. This study suggests a new mechanism of destabilization and aggregation. We also present an assay which recapitulates the proteotoxicity of these variants with a simplified and convenient model. Mechanisms and methods here reported will pave the way for the discovery and development of novel therapeutic strategies.

**INTRODUCTION**

Gelsolin (*GSN*) is a multifunctional regulatory protein responsible for the assembly, disassembly and scavenging of actin filaments through its severing and capping activities (1, 2). The protein is organized into six homologous domains (G1-G6), sharing the same *GSN*-like fold (3), each of which hosts at least one $Ca^{2+}$-binding site. In a $Ca^{2+}$-free environment, the six domains are closely packed together. Calcium binding induces both subtle local and large global conformational changes (3–6), where actin binding surfaces of G2, G1 and G4 become solvent-exposed. The crystal structure of the activated C-terminal half of GSN with/without actin (4, 7), showed a structurally conserved G4:G5 interface, suggesting that the interaction between the two domains is relevant for the inactive as well as the activated conformations of the protein.

Single point mutations of *GSN* gene are responsible for a systemic form of amyloidosis called Familial amyloidosis Finnish type (FAF), also known as AGel amyloidosis (AGel). This autosomal-dominant disease, first reported by Jouko Meretoja in 1969 (8), is characterized by the progressive deposition of amyloid fibrils in different organs and tissues (9–16).

Currently no pharmacological treatment blocking or slowing down any form of AGel is available, and only symptomatic treatments are being offered to improve the overall quality of life. AGel patients undertake several surgeries, transplants and other invasive and expensive medical procedures, which do not remove the source of toxicity (17)(18).

The most common forms of AGel are caused by substitutions in G2, including D187N/Y mutations which result in systemic progressive deposition of amyloids (9, 10), and N184K and G167R mutations, associated with kidney-localized aggregates (19–21).

The molecular mechanisms underlying the D187N/Y and N184K amyloidosis is well understood. These substitutions compromise G2 calcium binding site, leading to an overall destabilization of the domain (6, 22–24)(25, 26), with the exposure of an otherwise buried peptide, which is aberrantly cleaved by furin in the Golgi (27). The major product of furin activity, the C68 fragment, becomes a substrate of matrix metalloproteinases (MPP) and the proteolytic cascade eventually leads to the production of two aggregation-prone peptides of 5 and 8 kDa (28). The mechanism underlying G167R aggregation is instead yet to be fully elucidated: although this variant shows local destabilization and susceptibility to furin proteolysis similar to the others, this mutation also promotes domain-swapped oligomerization of the protein (26, 29, 30).

More recently, several novel pathological variants were reported (31–36), and among them A551P, E553K, and M517R cluster at the interface between G4 and G5. Little is known about the pathological mechanisms underlying these recently-identified AGel forms. Noteworthy, being furin a sequence-specific protease, G4 and G5 domains do not harbor any putative site potentially recognized by this protease.

Possible furin-alternative processes have been already suggested for a sporadic form of AGel associated with the deposition of wild-type (WT) GSN, which is resistant to furin proteolysis (37). *Ex vivo* mass spectrometry analyses of amyloid deposits from AGel patients carrying the G167R (20), A551P (31, 38) and unidentified mutations (19) revealed the presence of fragments not corresponding to the 5 and 8 kDa product of furin and MMP proteolysis. It is not yet known whether these peptides originate by alternative proteolytic patterns or by the fragmentation of full-length GSN during the analysis.

The three novel pathological variants, A551P, E553K and M517R, are ideal candidate systems to gain insight into alternative amyloidogenic pathways. Awaiting more clinical and histopathological findings that would reveal the nature of the aggregated material, we here performed a molecular profiling of the variants, evaluating their folding stability, aggregation and protetoxic potential and the impact of the substitutions on structure and dynamics of the protein.

# RESULTS

**G4:G5 variants do not increase GSN susceptibility to furin nor impair its physiological activity**

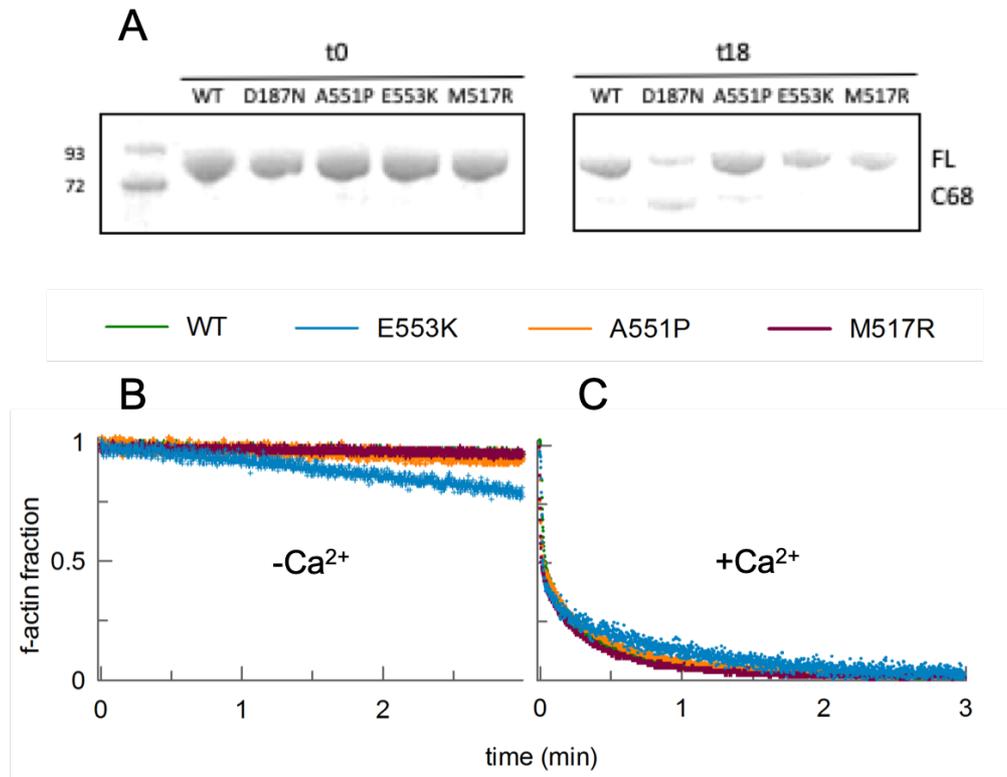

**Figure 1: Susceptibility to furin proteolysis and actin-severing activity. A)** WT, D187N, A551P, E553K, M517R GSN were analyzed by SDS-PAGE before (t0) and 18 h after incubation (t18) with furin. Full length (FL) and C68 fragment are indicated. **B-C)** Actin-severing activity of WT and mutated GSN was evaluated in the absence and in the presence of saturating $Ca^{2+}$ (**B** and **C**, respectively) by measuring the fluorescence intensity of pyrene-labelled f-actin over time.

All the amyloidogenic GSN variants characterized so far in domain G2 showed an increased susceptibility to furin-dependent proteolysis *in vitro (24–26, 29)*. To investigate whether A551P, E553K and M517R substitutions affect the proteolytic process, we first evaluated their sensitivity to furin using the WT and D187N GSN as controls. As expected, only the D187N variant produced a significant amount of the C68 fragment, the larger product of furin proteolysis. A faint band, compatible with a molecular weight slightly greater than the C68 fragment, was observed for A551P and E553K, as well as the WT GSN, suggesting that they are either minimally cleaved by furin, or subjected to an unspecific proteolytic event, as previously reported (24)(Figure 1A).

The PROSPER algorithm, a machine-learning based predictor of protease cleavage sites, excluded that the single point substitutions of A551, E553 and M517 could generate novel protease cleavage sites (39).

We tested whether the mutations affect GSN physiological function, *i.e.* its ability to bind and sever actin filaments in a pyrene-labeled fluorometric assay. Under $Ca^{2+}$-free conditions, where GSN adopts the closed and compact conformation, the actin binding and severing activity were negligible (0.62 ± 0.01 $h^{-1}$ for WT) and no pathological mutant analysed so far showed a different behaviour (4, 26). Same for the A551P and M517R variants (1.10 ± 0.01 and 0.78 ± 0.01 $h^{-1}$, respectively) but not for E553K mutant which surprisingly showed a 6-fold higher severing activity (3.80 ± 0.01 $h^{-1}$) even in the absence of $Ca^{2+}$ (Figure 1B). Although such $Ca^{2+}$-independent activity is probably not sufficiently high to be relevant in a physiological environment, it does suggest that E553K substitution compromises the compact inactive state of the protein. In the presence of $Ca^{2+}$, all the tested variants are endowed with high severing efficiency similar to that of WT protein (Figure 1C) suggesting that no loss-of-function is associated with these GSN substitutions.

**G4:G5 mutations affect both the stability and the unfolding behaviour of GSN**

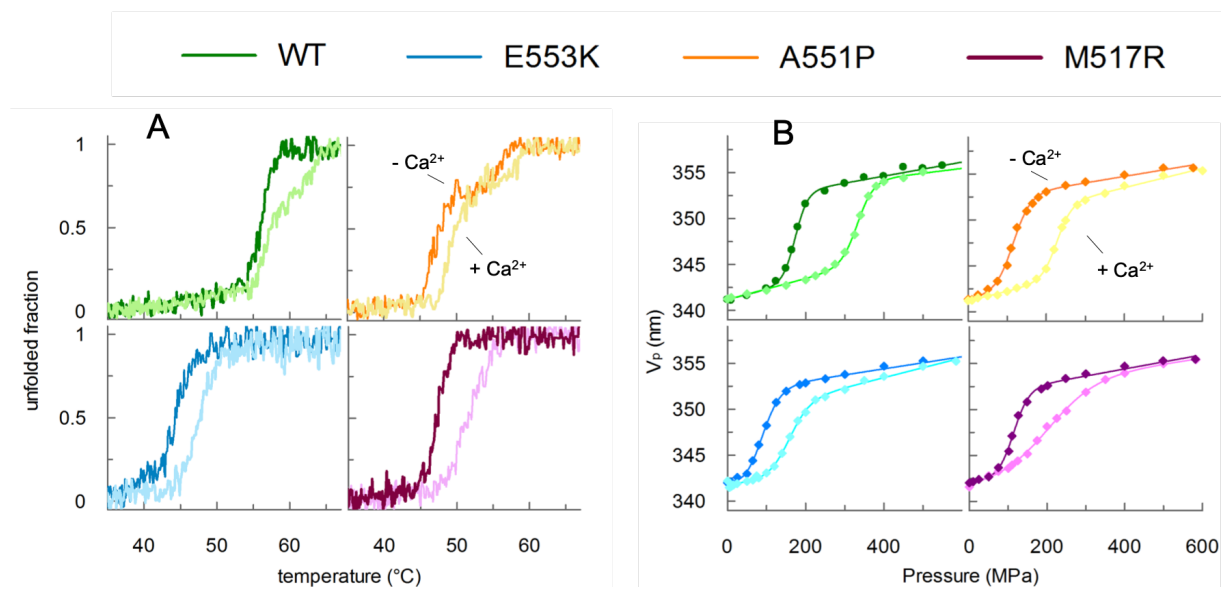

**Figure 2: Impact of the mutations on temperature and pressure stability. A)** Thermal denaturation of WT GSN and A551P, E553K, and M517R variants analysed by circular dichroism spectroscopy, both in the absence (darker shade) and in the presence (lighter colour) of saturating $Ca^{2+}$. **B)** Pressure-induced denaturation at 20 °C followed by the change in tryptophan fluorescence emission (same color code). $V_p$ is the center of spectral mass (nm) of tryptophan emission. Lines show the fit of the data with a two-state equation with sloping baselines (40).

| | $T_{on}$ (°C) | | | $P_m$ (MPa) | | $\Delta V_0$ (mL/mol) | |
|---|---|---|---|---|---|---|---|
| | $-Ca^{2+}$ | $+Ca^{2+}$ | | $-Ca^{2+}$ | $+Ca^{2+}$ | $-Ca^{2+}$ | $+Ca^{2+}$ |
| *WT* | 53.0 ± 2.7 | 53.8 ± 3.0 | | 174 ± 18 | 333 ± 21 | -152 ± 16 | -118 ± 8 |
| *A551P* | 44.2 ± 1.8 | 46.4 ± 3.8 | | 116 ± 16 | 226 ± 25 | -109 ± 9 | -125 ± 7 |
| *E553K* | 39.9 ± 2.3 | 46.8 ± 3.7 | | 86 ± 10 | 157 ± 14 | -103 ± 8 | -91.3 ± 9 |
| *M517R* | 44.3 ± 3.7 | 49.6 ± 1.7 | | 121 ± 16 | 214 ± 13 | -122 ± 17 | -34 ± 1 |

**Table 1: Stability parameters for WT and mutated GSN.** Mutation-dependent destabilization was evaluated either in the absence or presence of $Ca^{2+}$ in thermal denaturation experiments followed by circular dichroism in the far-UV region and the temperature of onset of denaturation is reported ($T_{on}$). Protein stability was measured under the same experimental conditions, in pressure denaturation studies followed by intrinsic fluorescence at 20 °C. The pressure of midpoint denaturation ($P_m$) and the difference of volume of the system, unfolded-folded, ($\Delta V_0$) are tabulated. Standard errors of the fitted parameters are reported.

To investigate the effect of the G4 and G5 mutations on the stability of GSN, the three variants were subjected to temperature and pressure denaturation studies following circular dichroism (CD) and tryptophan fluorescence signals, respectively, and the analysis was repeated for the two functional states of GSN, active and inactive. Irrespective of the tested conditions, both techniques showed that G4:G5 substitutions cause a significant destabilization of the protein fold.

Although irreversible, thermal denaturation of GSN was successfully used to characterize other G2-linked pathological variants (D187N, N184K and G167R) (26, 29, 30). In the absence of $Ca^{2+}$, WT GSN displays a prototypical single transition behaviour which is consistent with the compact and nearly globular conformation of the protein (Figure 2A). In the presence of the ion, WT GSN unwinds, many inter-domain contacts are modified or lost and the denaturation occurs via a multi-state process (26, 29). Surprisingly, all the variants also showed different qualitative behaviors, suggesting a complex role of the mutated residues. The A551P mutation led to a stepwise denaturation even in the absence of $Ca^{2+}$, while a single transition was always observed for the E553K and M517R variants.

Because of the differences in melting behavior, to quantify the mutation-dependent destabilization, we considered the temperature associated with the first denaturation events (temperature of onset, $T_{on}$). $T_{on}$ values measured for mutated GSN were lower than those of WT protein (Table 1). In the absence of $Ca^{2+}$, $\Delta T_{on}$ (mutant-WT) is -9 °C for A551P, -13 °C for E553K, and -9 °C for M517R. The addition of $Ca^{2+}$ had little effect on the $T_{on}$ of WT and A551P proteins, while the extent of destabilization is conformation-specific for E553K and M517R. E553K appeared to be the most unstable of all the variants, with a $T_{on}$ of 39.9 ± 2.3 °C in the absence of $Ca^{2+}$ (in agreement with the actin-severing activity, Figure 1B).

Thermal stability of the G5 variants was also evaluated on the isolated domain. The A551P substitution (Figure S1) induced only a sligh destabilization of the isolated domain with respect to G5-WT ($T_m$ of 63.4/55.8 °C with/without $Ca^{2+}$), with a $\Delta T_m$ of -4.4 °C in $Ca^{2+}$ and -5.1 °C without $Ca^{2+}$, while the E553K substitution strongly destabilized G5-E553K, with a $\Delta T_m$ with respect to G5-WT of -14.0 °C and -19.3 °C, respectively.

Impact of the mutations on the stability of full length GSN proteins was also evaluated by pressure-induced denaturation followed by intrinsic fluorescence emission. Trp fluorescence monitors, in addition to the folding state of the individual domains, the loss of inter-domain contacts and is thus informative of the conformational state of the protein. The denaturation can be modelled by a standard two-state equation with sloping baselines (Figure 2B), a pattern already observed in chemical denaturation studies on WT and D187N/Y GSN (23), and interpreted as the result of highly dynamic folded and unfolded states. Pm and $\Delta V_0$ values (reported in Table 1), defined as the pressure of the transition midpoint and the volume change between folded and unfolded states, respectively, are used to quantify the mutation-dependent destabilization.

The Pm parameters showed a good agreement with the thermal denaturation Ton and confirmed the stability scale: WT>A551P≥M517R>E553K. All the variants showed a reduction of the differential stability between the $Ca^{2+}$ bound/unbound states with respect to that of WT ($\Delta P_m$ =155 MPa), which is modest for A551P ($\Delta P_m$ =115 MPa) and more pronounced for E553K and M517R (85 and 73 MPa, respectively).

Assuming similar unfolding states for all the variants and in all conditions, higher net $\Delta V_0$ values suggest a more compact folded molecule. Indeed, net $\Delta V_0$ is higher for the globular inactive GSN (-152 mL/mol) with respect to its dynamic and extended $Ca^{2+}$-bound conformation (-119 mL/mol). In comparison with WT, all mutations caused a 20-30% reduction of net $\Delta V_0$ under $Ca^{2+}$-free conditions, in particular for A551P and E553K (-109 and -103 mL/mol, respectively), while the impact on this parameter in the presence of the ion is of more difficult interpretation. In particular, M517R shows a very low value (-34 mL/mol) but likely as a result of a loss of cooperativity of the process as suggested by the gradual denaturation curve.

**G4:G5 mutations grant aggregation propensity to the unproteolysed full length protein**

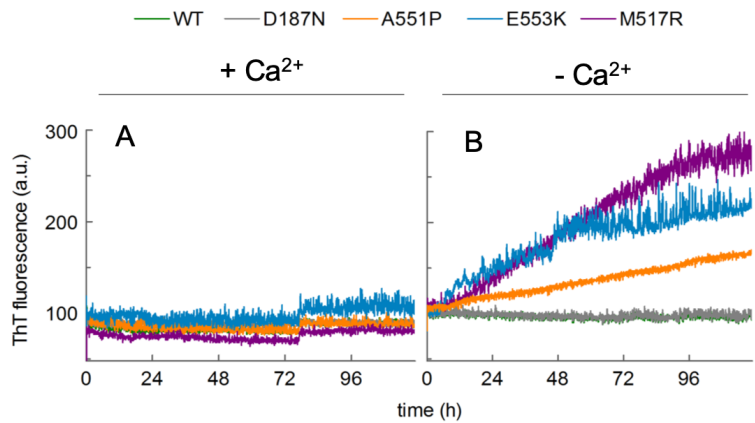

**Figure 3: Aggregation propensity of WT and mutated GSN.** Propensity to form aggregates was evaluated by fluorimetric ThT assay on 10 µM GSN samples. Experiments were performed both in the presence (**A**) and in the absence (**B**) of saturating $Ca^{2+}$.

To dissect the determinants and mechanisms underlying the pathological deposition of the A551P, E553K and M517R variants, we investigated their aggregation propensity *in vitro* by Thioflavin T (ThT) assay. This fluorogenic compound is specific for amyloid-like structures both as soluble prefibrillar aggregates and mature fibers and, because of this property, it is widely used in diagnostics and kinetics studies.

The aggregation propensity of the G4:G5 variants, WT and D187N proteins, in the presence and absence of $Ca^{2+}$, was evaluated at 37 °C under stirring conditions (Figure 3). In the presence of $Ca^{2+}$, no increase of ThT signal was recorded for WT and mutated GSNs over a 5-day incubation (Figure 3A). A similar behavior is observed in $Ca^{2+}$-free conditions for WT and the G2-linked D187N variant (Figure 3B), which were already reported not to aggregate in their unproteolysed form and under nearly physiological conditions (28). On the contrary, for the A551P, E553K and M517R variants we observed an increase of the ThT signal indicative of their propensity to aggregate. This is particularly evident for E553K and M517R.

A sequence-based amyloigenicity analysis performed with Amylpred2 (41) did identify several potential aggregation-prone sequences in WT GSN, including some close to the G4:G5 interface (Table S1). In addition to G2, fragmentation or misfolding of G4 and G5 domains may lead to GSN deposition.

**Both sequence and structure contribute to G4:G5 mutants (proteo)toxicity**

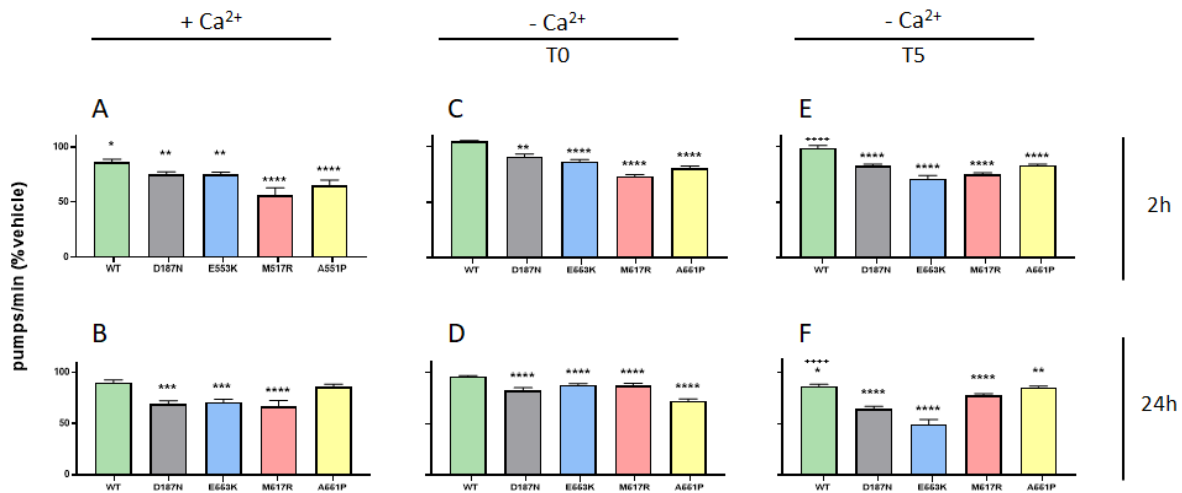

**Figure 4:** *In vivo* toxicity of full length GSN is related to the protein sequence and conformation. Proteotoxicity of GSN samples and control (vehicle) was evaluated in the presence of $Ca^{2+}$ (**A, B**) and in its absence (**C, D, E, F**), also after a 5-day incubation at 37 °C (**E, F**). The pharyngeal activity was determined 2 h (**A, C, E**) and 24 h (**B, D, F**) after feeding GSN variants by determining the number of pharyngeal bulb contraction (pumps/min). Data are mean (**A, B, C, D**) ± SE (N=30/group). *p< 0.05, **p< 0.01, ***p< 0.001 and ****p<0.0001 *vs* Vehicle, according to one-way ANOVA and Bonferroni *post hoc* test; and (**E, F**) ± SE (N= 10/group). *p<0.05, **p<0.005 and **** p< 0.0001 vs Vehicle and ++++ p< 0.0001 *vs* mutated GSN.

To characterize the proteotoxic potential of the novel G4:G5 variants *in vivo*, we applied a *C. elegans*-based toxicity assay. *C. elegans* is able to specifically react to the toxic assemblies of amyloidogenic proteins, reducing the function of its pharynx, defined as *pumping rate* (42–45). Employing this approach, we had already observed a good correlation between the mutation-induced destabilization and amyloidogenicity of isolated G2 domains (24). Full length proteins were administered to worms in the presence or absence of $Ca^{2+}$ (Figure 4A-D). In the absence of $Ca^{2+}$, experiments were also repeated 5 days after an incubation under conditions similar to those of the aggregation experiments (Figure 4E-F). The pumping rate of nematodes was measured 2 h and 24 h after the administration to evaluate the transient and permanent toxic effect, respectively.

In the presence of Ca$^{2+}$, all tested GSN variants caused a reduction of the pharyngeal activity, ranging from 13% (WT) to 37% (M517R) (Figure 4A). The effect lasted overtime except for the WT and the A551P protein, which reverted to vehicle-levels 24h after feeding (Figure 4B). A less toxic effect was overall observed when the mutated proteins were administered to worms in the absence of Ca$^{2+}$ and the WT protein was not recognized as toxic under these conditions. The reduction of the pharyngeal pumping at 2 h ranges 10-27%, following the same toxicity scale as in the presence of Ca$^{2+}$ (Figure 4C). Under this experimental condition, the A551P variant became able to induce a permanent dysfunction (Figure 4D). Aging of D187N, M517R and A551P proteins, resulted in a similar reduction of the pharyngeal function scored 2 hours after the treatment, whereas E553K was much more toxic after the incubation (Figure 4E). The toxicity of the aged form of D187N, E553K and M517R, but not A551P, increased over time indicating their strong ability to permanently affect the pharyngeal function of worms (Figure 4F). The data obtained suggest that the various mutations induce a destabilization of the proteins over time with different toxic properties.

Since thermal stability and time-dependent experiments showed abundant amorphous precipitation of the most destabilized variants, we also evaluated whether protein denaturation by flash-heating may affect their proteotoxicity (Figure S2). Denatured proteins lose their ability to impair the pharyngeal function of worms demonstrating that *C. elegans* specifically recognizes the mutation-dependent toxicity of the folded proteins, whether in the open, closed or aggregated conformations.

**The mutations affect structure and dynamics of the G4-G5 domains**

To dissect the structural determinants of the mutation-dependent destabilization, aggregation propensity and proteotoxicity we performed X-ray crystallography and molecular dynamics studies.

While the elusive Ca$^{2+}$-bound conformation of full length GSN has never been obtained, crystallogenesis of the Ca$^{2+}$-free protein is generally reproducible. All G2-linked mutants and the WT protein have been previously crystallized in similar conditions leading to easy to compare (same crystal lattice and similar quality of diffraction data) atomic models (3, 26, 30). Despite multiple trials, E553K and M517R proteins only grew poorly diffracting crystals, underlying once more the high destabilization of these substitutions.

Crystals of A551P in the Ca$^{2+}$-free conformation, instead, diffracted to 3.0 Å. Quality of the diffraction data and of the model is comparable to the previously published structures (Table

S2). Most of the residues belonging to the six homologous domains could be unambiguously traced in the electron density with the exception of some flexible linkers and few stretches of G5. In this domain a noisy electron density did not allow accurate placement of side chains and sufficient level of details for fine orientation even of the backbone, which forces to model a near-to-ideal structure similar to that of WT (Figure S3). Analysis of the normalized B factor/residue becomes a valuable tool to quantify the increased dynamics of the G4G5 portion of the protein (Figure 5A). The G5 domain is characterized by a higher conformational flexibility even in the WT protein. A551P substitution did not significantly increase the dynamics of the mutation site but rather that of the already flexible stretches, residues 450-460, 525-540 and 570-575.

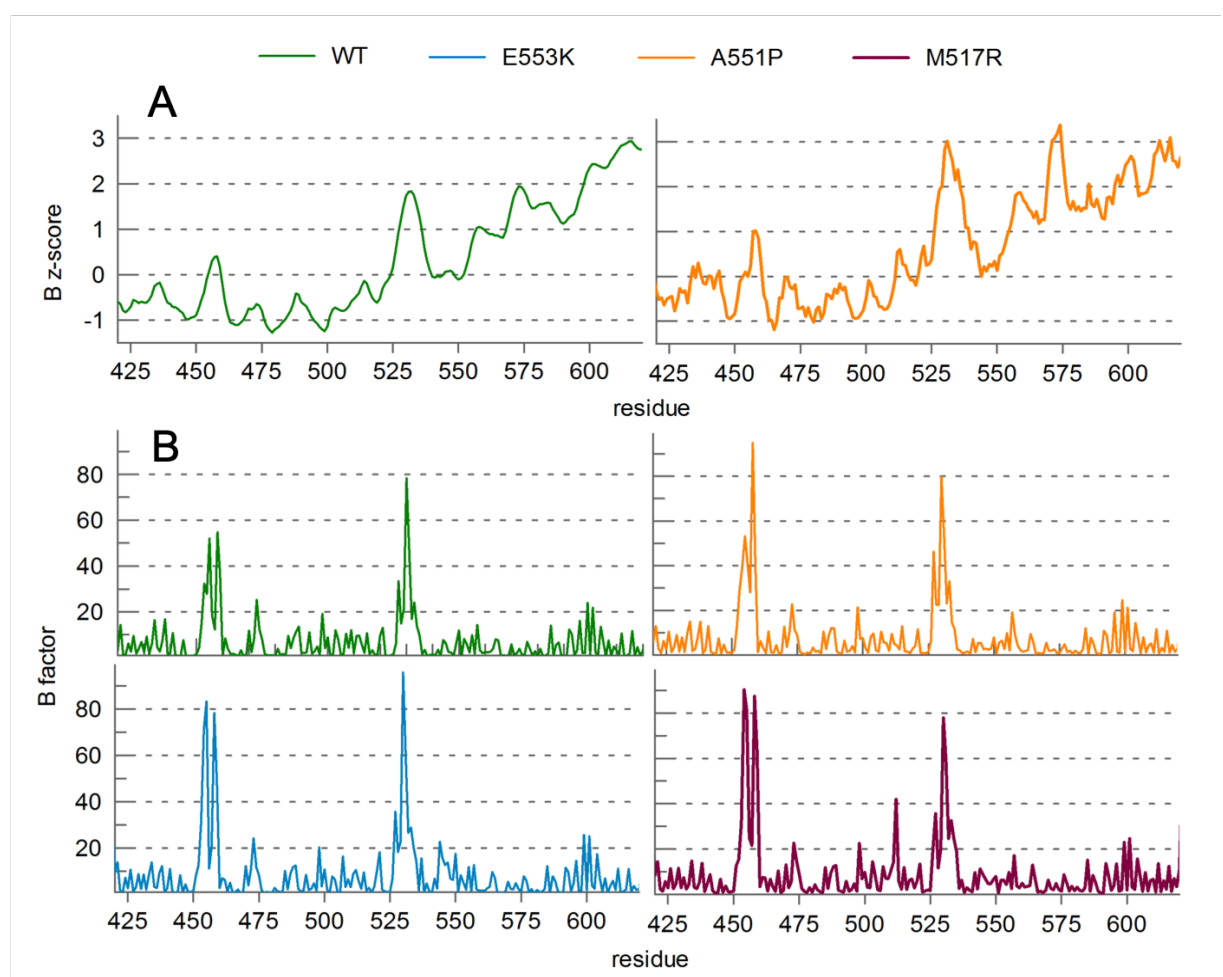

**Figure 5: Mutation-induced local conformational flexibility.** Displacement of atoms of the backbone of the G4-G5 region are reported **(A)** as the temperature factors of the atomic models obtained by crystallographic data for the WT protein (PDB id: 3FFN (3)) and A551P (this study). Both are expressed as Wilson B-factors, normalized in the latter analysis (B z-score); and **(B)** as the fluctuations with respect to their initial position throughout molecular simulations of a G4-G5 construct.

Unrestrained explicit-solvent molecular dynamics simulations of the G4-G5 pairs (residues 412-630) were thus performed to explain the role of the substitutions in the destabilization of the domains. The A551P, E553K and M517R variants were subjected to triplicate runs, each 1 μs long, which is sufficient to observe mutation-induced local rearrangements but not long enough to sample large-scale conformational changes, such as those required for amyloid-like aggregation.

As a first analysis, we identified the most mobile regions of the structures computing the fluctuation of each residue with respect to their initial position over the simulation time, plotted as B factor in Figure 5B. In accordance with the crystallographic data, the areas of greater conformational flexibility, common to all the analysed variants, are the linker connecting the G4 and G5 (residues 520-530) and the 450-460 loop. Interestingly, E553K showed higher flexibility also in the region between residues 540 and 550, which lies at the interface between the G4 and G5 domains. This flexibility may explain the observed actin severin activity, thus suggesting that this interface might be partially opened and bind actin. A strong local effect is observed for the M517R variant with a B factor peak corresponding to the mutation site.

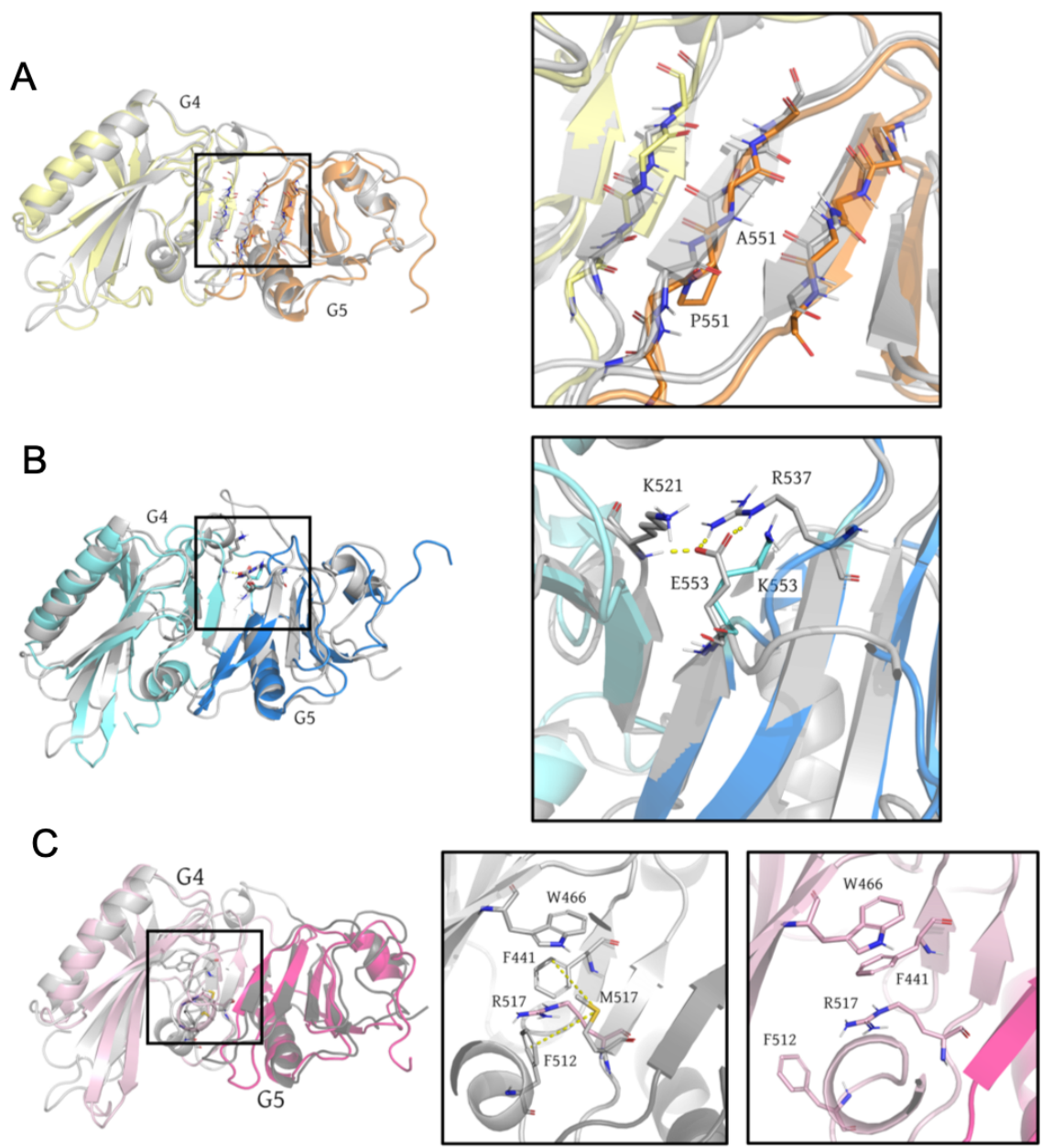

**Figure 6: Structural assessment of the impact of the G4 -G5 mutations on the interdomain arrangement of GSN**. Final frames of MD simulation of (**A**) A551P, (**B**) E553K and (**C**) M517R are represented as cartoon and colored with a light/darker (domain G4/G5) shade of yellow, blue and magenta, respectively; and are superimposed on the final frame of the WT simulation, in grey. Images on the right show details of the respective mutation site: mutated and interacting residues or those displaying significant displacements are shown as sticks and labeled.

To rationalize the structural impact of each point mutation we superimposed the starting protein configurations and those after 1 μs of simulation and computed the corresponding root-mean squared distance (RMSD), equal to 1.85, 2.10, 2.30 and 2.54 Å for the WT, A551P, E553K and M517R proteins, respectively (Figure 6). The comparison of these RMSD values for the three variants suggests an increasing dynamicity of the overall G4-G5 portion which is 13% to 37% higher than in the WT.

Both in the $Ca^{2+}$-bound and $Ca^{2+}$-free forms of WT GSN (3, 7), G4 and G5 domains are paired within an interface consisting of two parallel β strands (residues 517-520 and 549-553, respectively). The two domains establish a tight packing through an extended H-bond network that contributes to global folding and stability of the protein. In the A551P mutant we observed a loss of the planarity of both edge β-strands of G4 and G5 and of some helical portions. Proline residues are β breakers and impair strand-strand interactions (46). The inter-strand bond between A551 and M517 was lost because the closure of amino N in a pyrrolidine ring exhausts the electronic demand for N (Figure 6A). In the WT protein, E553 is involved in salt bridges with the guanidine of residue R537 and the amide group of K251 which confers high conformational stability (Figure 6B). The E to K substitution caused local rearrangements due to charge repulsion, which led to an increased local flexibility. M517R was the variant characterised by the highest RMSD and showed a strong destabilization of the G4:G5 interface. In the WT protein, M517 extends its side chain towards a sulphur-aromatic motif (S-π, (47)) which includes residues F441, W466, M509 and F512 in the core of G4 (Figure 6C). The M517R mutation caused important upheavals of the S-π motif, as it introduces a charged and bulky residue. In addition to the loss of the π-S bond, the long side chain of R discouraged stabilizing interactions with neighbouring residues.

To support the qualitative information about the destabilization of the G4:G5 interface induced by the mutations, we performed a statistical evaluation of the strength of the hydrogen bonds network between the backbone of the strands at the G4:G5 interface, measuring the occupancy of each inter-strand interaction along the simulations. This analysis, whose details are reported in the supplementary (Table S3 and Figure S4) highlighted a loss of H-bond connectivity of the interface for the three variants, namely equal to -0.2, -1.19 and -1.25 bonds for M517R, E553K and A551P, respectively.

**DISCUSSION**

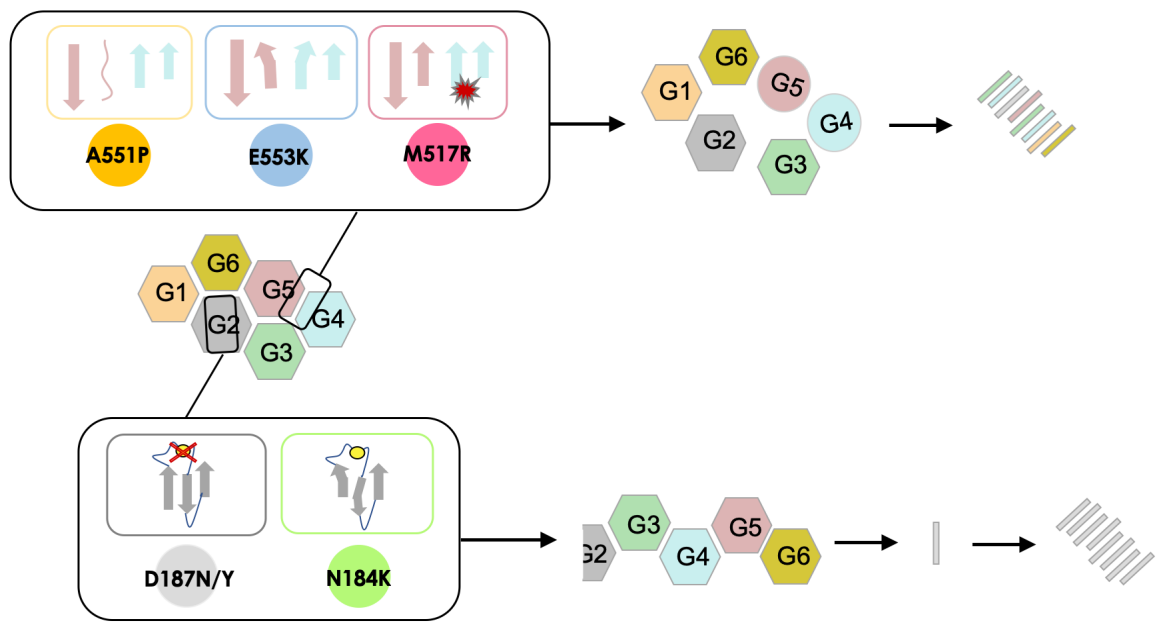

**Figure 7: Different mechanisms of destabilization may trigger alternative amyloidogenic pathways.** Graphical description of the impact of mutations on the hosting domain. Previous studies showed that D187N/Y impairs $Ca^{2+}$-binding in G2 (23, 24, 48), with the ion having a structural role, whereas N184K causes similar destabilization due to a loss of connections without affecting the capacity to bind the ion (25, 26). Both mutations lead to the exposure of a proteolysis-sensitive site and trigger the canonical furin-dependent aggregation mechanism. Despite showing local destabilization and susceptibility to furin proteolysis, the mechanism underlying G167R aggregation is instead yet to be fully elucidated (29). The three mutations focus of this study (A551P, E553K and M517R) are not cleaved by furin, yet they disturb the interface between G4 and G5, where the substitutions cluster. Three different mechanisms of destabilization were identified: distortion of the edge strand, charge repulsion, destabilization of the G4 fold, for the A551P, E553K and M517R variant, respectively.

An ever increasing number of amyloidogenic variants of GSN have been identified over the years as responsible for AGel, but only those hosting the substitutions in the G2 have been so far characterized at a molecular level. Little is known about the recently identified A551P, E553K and M517R mutations hosted in the interface between G4 and G5 domains, whose aggregation process likely follows an alternative pathway. Noteworthy, G4:G5 interface does not harbor any putative site potentially recognized by furin (the protease is sequence-specific), nor the mutations should impact the stability or the susceptibility to proteolysis of G2, thus representing an ideal candidate system to explore alternative amyloidogenic pathways.

After excluding the furin cleavage as triggering factor for the aggregation of the three pathological variants and any effect on the physiological activity of GSN, we focused on the intrinsic stability of the proteins, their aggregation propensity and toxicity, and a structural perspective of the effect of the substitutions on the protein (summarized in Figure S5, also in comparison with the already characterized G2-linked substitutions). These findings are a step forward toward the identification of a link between mutation and clinical phenotype, which is at the moment limited also by the absence of a proper prognostic biomarker for AGel.

The A551P variant was identified in a patient presenting, in addition to AGel, another rare amyloidosis caused by a mutant of the transthyretin protein (ATTR-V122I) (31, 38). The patient did not show any clinical signs typical of AGel and overall A551P deposition appeared to be responsible for a mild phenotype. A551P is also the only G4:G5 variant whose available clinical findings hint at a novel underlying molecular mechanism, since *ex vivo* analysis of the deposits identified non-canonical peptides either products of alternative proteolytic pathways or of the deposition of the full length protein. As expected, the impact of A551P mutation on the molecular features of GSN is mild (Figure S5), without large rearrangements and modest destabilization of the protein. The A551 residue is hosted by the edge β-strand of G5, the substitution with a proline causes the loss of one H-bond and a distortion of the strand itself, which loses its planarity, and weakens connectivity of the interface (Figure 6A and Table S3). Yet such minor rearrangements, in particular under $Ca^{2+}$-free conditions, are sufficient to disturb the overall protein architecture and metastability basins, and expose aggregation-prone stretches of the protein.

The S-π motif, of which M517 is part of, appears to be a major determinant of G4 stability. The M517R substitution destroys the cluster of hydrophobic residues deep in the domain core, due to the insertion of a charged and bulky residue. As a result, M517R is the variant showing the most diverse flexibility spectrum (Figure S5D), with a B factor peak in the mutation area. However, such increased conformational flexibility only results in a modest destabilization, in the A551P range (Figure S5C), as evaluated by thermal and pressure denaturations of the full length protein. Although this mutation has a strong impact on protein structure and dynamics, the effect is rather localized and the slight destabilization of the overall architecture of the protein is likely only a consequence of the misfolding of the domain hosting the mutation. We observed the same mechanism for other, previously characterised, GSN mutations (24–26). The M517R mutant is the one showing highest propensity to form aggregates, and also the highest transient toxicity in the *C. elegans* assay (Figure S5E-F). These findings suggest that future studies should focus on the domain hosting this mutation, looking for aggregation prone sequences or cleavage sites of yet-to-be-identified proteases.

At the time the manuscript was being finalised, a novel pathological variant, namely W466R, was identified in the same S-π motif in G4 (Figure 6C) (49). This variant causes a clinical phenotype similar to M517R, E553K and the classical D187N/Y mutations, including corneal lattice dystrophy, cutis laxa and peripheral neuropathy. The substitution introduces a positive charge in an otherwise hydrophobic region, suggesting that a similar destabilization mechanism underlies both M517R and W466R variants.

The strongest molecular phenotype was observed for the E553K variant (Figure S5). The one characterized by the lowest thermal and pressure stability both when measured on the isolated domain and on the full length protein, suggesting that the substitution disturbs the folding of the G5 domain and its interaction with G4. The mutated protein adopts a partially opened, i.e. active and able to bind actin, conformation even in the absence of calcium. With onset of denaturations near to physiological conditions, this labile variant was challenging to characterize. Abundant amorphous deposition was observed prevalently in all time course experiments but these insoluble aggregates were shown not to be toxic. The variant is also prone to form ThT-positive and toxic soluble aggregates, similarly to M517R, with whom shares a similar clinical picture, but aggravated in E553K patients also by a cardiac involvement (32, 33). In the E553K protein, a basic residue substitutes an acidic one in an area already crowded with positive charges, leading to intra- (G5) and inter-domain (G4:G5) charge repulsion. As a consequence, the two domains are pushed away and local connectivity is significantly impaired. A good correlation between propensity to aggregate and toxicity was observed for the E553K variant, for which we observed the highest permanent impairment on pharyngeal pumping at T5. Whether this is due to quantitative or qualitative features of the aggregates remains to be evaluated.

In conclusion, we have identified a novel hotspot of GSN instability and pathogenicity: the interface between domain G4 and G5. Another intradomain interface (G2:G3) was recently shown to be relevant for GSN amyloidogenicity (30). Researchers were in fact able to reproduce the amyloidogenic pathway underlying the deposition of D187N/Y GSN by engineering mutations that impair the interaction between domain G2 and G3. We also demonstrated that different mechanisms of destabilization, namely strand distortion, charge repulsion and steric hindrance, can all cause local destabilization or loss of connectivity sufficient to relax GSN global architecture (Figure 7). The reorganization of the domains, in particular in the absence of calcium ions, that stabilize the individual domains, leads to the exposure of stretches of the protein prone to engage in aberrant interactions. This novel mechanism of GSN pathological aggregation is here proposed to stimulate more clinical research, in particular on the chemical nature of patients' deposit. Although the larger number of diagnoses, including those caused by new variants, suggest an increased awareness of the disease, AGel is still understudied.

## MATERIALS & METHODS

### Protein production

The A551P, E553K and M517R variants in the full length gelsolin and the isolated G5 domain were produced by site-directed mutagenesis, using the WT construct as a template and the Q5® Site-Directed Mutagenesis Kit (New England BioLabs). Primers were designed using the manufacturer's software (nebasechanger.neb.com). Wild-type (WT) full length human GSN and the three variants were expressed as heterologous proteins in E. coli cells and purified following the protocols reported in (26, 29), while WT G5 domain and its A551P and E553K variants were produced using the protocols for the isolated G2 reported in (25, 29).

### Furin proteolysis assays

Furin cleavage assays were performed as reported in (24) using 4 U of commercial furin enzyme (BioLabs) and 1 mg/ml of the WT, D187N, A551P, E553K e M517R variants in 20 mM MES, pH 6.5, 100 mM NaCl, 1 mM $CaCl_2$. The final volume of each reaction mix was 60 µl and the reaction was performed at 37 °C. Proteolysis was monitored by SDS-PAGE.

### Fluorimetric actin severing assay

Pyrene-labeled rabbit skeletal muscle globular actin (G-actin; Cytoskeleton, Inc. (Denver, CO, USA)) was used to evaluate the severing activity of the pathological variants of gelsolin. Preparation of the solutions, actin manipulation and conversion of G-actin to filamentous actin (F-actin) were performed as reported elsewhere (50). Measurements were performed at 20 °C with a Cary Eclipse fluorimeter (Agilent Technologies, USA), with the following settings: excitation and emission wavelength/slit of 365/5 and 407/5 nm, respectively; averaging time 0.1 s. A 3 ml cuvette was used, hosting a stirring bar for continuous agitation of the reaction mixture. 400 µl of a 4 µM F-actin solution were incubated in the cuvette until stabilization of the fluorescence signal, then 2 µl of 50 µM GSN were added (0.25 µM final concentration). Once the signal was stable again, the severing reaction was started by adding 2 µl of a 1 M $CaCl_2$ solution (final free $Ca^{2+}$ concentration > 1 mM). Data were normalized based on initial and end-point fluorescence measured in the presence of $Ca^{2+}$. For the $Ca^{2+}$-free assays, measurement is started right upon addition of the proteins and depolymerization rate calculated by linear fitting over 3 min.

**Denaturation monitored by circular dichroism and fluorescence emission.**

Thermal stability of the GSN variants was evaluated as previously reported (26). Briefly, proteins were diluted to 0.2 mg/ml in 20 mM HEPES, pH 7.4, 100 mM NaCl and either 1 mM EDTA or 1 mM $CaCl_2$. Loss of protein secondary structures was monitored following circular dichroism at 218 nm during a 20 to 95 °C temperature ramp (1 °C/min). Temperature of onset ($T_{on}$) of denaturation for the full length proteins was calculated by fitting the linear portion of the curve and at the beginning of the denaturation. For the isolated domains which showed a standard two-state behaviour, the melting temperature ($T_m$) was instead calculated as reported (29).

Fluorescence spectra of wild type and mutants GSN were monitored at equilibrium as a function of pressure, from 0.1 to 600 MPa, at 20 °C, on a homemade apparatus using a pulsed excitation ($\lambda_{ex}$=292 nm) (51). Fluorescence emission in the spectral range 300–430 nm was monitored by a back-illuminated 1340x400 pixels CCD camera (Princeton Instruments Spec-10:400B (XTE) Roper Scientific, Trenton, NJ), cooled to -60 °C. The reversibility of the changes induced by pressure was checked at the end of each pressure cycle. Pressure effects were not promptly and completely reversible for all the proteins analyzed: after 12 hours the recovery of the fluorescence characteristics was 70-80% for WT and less (about 50%) for the mutants, both in the absence (1 mM EDTA) and in the presence of $Ca^{2+}$ (50 mM): The delayed recovery suggests that after decompression there was a slow and continuous reorganization of the loose structure. Folding/unfolding transitions of multidomain proteins are usually characterized by the presence of a few partially folded equilibrium intermediates in which folded and unfolded domains coexist (52, 53). Usually these intermediate states are produced at a fast rate, and slowly evolve to the native one (52) At each pressure, the spectral changes in protein fluorescence emission were quantified by determining the center of spectral mass, defined as $v_P = (\Sigma v_i F_i)/F_T)$ where $F_i$ is the fluorescence intensity at the wavenumber $v_i$, and $F_T$ is the total fluorescence emitted from the protein. The fraction of unfolded protein was determined at each pressure from the displacement of $v_P$ (54). Pressure unfolding was analyzed following a two-state ($N \leftrightarrow U$) model (55), implemented with sloping baseline correction (23).

**Crystallization, structure solution and analysis**

Crystallization trials were performed using an Oryx-4 nanodispenser robot (Douglas Instrument) by using the sitting drop vapor-diffusion method. Experiments were carried out using A551P concentrated to 10 mg/ml (120 µM) in 20 mM HEPES, 100 mM NaCl, 1 mM EGTA, 1 mM EDTA, pH 7.4 at 20 °C. The best diffracting crystals appeared in 1.3 M ammonium sulfate, 100 mM Tris–HCl, 15% glycerol, pH 8.5. X-ray diffraction data of A551P were collected at beamline I04 at Diamond Light Source (Harwell Science and Innovation Campus in Oxfordshire). Data of A551P were processed using XDS (56) and scaled with AIMLESS (57). Structures were solved by molecular replacement with PHASER (58) using the WT gelsolin crystal structure (PDB ID 3FFN (3)) as a search model. *Phenix refine* (59) was used for the refinement of the structure while the manual model building was performed with Coot (60). The structure of full length A551P (two molecules in the asymmetric unit) was deposited with PDB ID 7P2B. Owing to a better quality of the electron density for chain A, this molecule is used for the analyses. Analysis of the structures was performed with PyMOL (Schrödinger; DeLano 2002), which was also used to prepare the figures. B-factors were normalized (Bz-score) and analyzed as reported in (61).

**Oligomerization kinetics monitored by ThT fluorescence**

Aggregation kinetics were followed using an *in situ* ThT fluorescence assay based on the increase of the fluorescence signal of ThT when bound to β sheet-rich structures (62). The different variants, at a final concentration of 10 µM, were incubated under continuous orbital shaking in 20 mM Hepes buffer pH 7.4, 100 mM NaCl and in the presence of either 1 mM $CaCl_2$ or 1 mM EDTA, at 37 °C in microplate wells (Microplate Corning 3881, 96-well, low-binding, Corning Inc. Life Sciences, Acton, MA) in the presence of 20 µM ThT and 0.02% $NaN_3$ (100 µl solution/well). ThT fluorescence was measured every 5 min using an F500 Infinity plate reader (Tecan Italia Srl, Cernusco Sul Naviglio, Italy). The dye was excited at 448 nm, and the emission was measured at 485 nm.

**Proteotoxicity studies on *C. elegans***

Bristol N2 strain was obtained from the *Caenorhabditis elegans* Genetic Center (CGC, University of Minnesota, Minneapolis, MN, USA) and propagated at 20 °C on solid Nematode Growth Medium (NGM) seeded with *E. coli* OP50 (CGC) for food. The effect of WT GSN and GSN carrying D187N, A551P, E553K or M517R mutation on pharyngeal behavior was evaluated as already described (24). Briefly, worms were incubated with 1.5 µg/ml of protein (100 worms/100 µl) in 20 mM HEPES solution containing 100 mM NaCl (Hepes solution) and 1 mM $CaCl_2$ or 150 µM EDTA, pH 7.4. Hydrogen peroxide (1 mM) was administered in dark conditions as a positive control (100 worms/100 µl). Control worms were fed 20 mM HEPES solution with 1 mM $CaCl_2$, pH 7.4 (100 worms/100 µl) only.

After 2 h on orbital shaking, worms were transferred onto fresh NGM plates seeded with OP50 *E. coli*. The pharyngeal pumping rate, measured by counting the number of times the terminal bulb of the pharynx contracted over a 1 min interval (pumps/min)., was scored 2 and 24 h later. In selected experiments *C. elegans* were fed 1.5 µg/ml GSN (100 µL/ 100 worms) previously incubated or not at 37 °C for 5 days in 20 mM Hepes solution containing 1 mM $CaCl_2$ or 150 µM EDTA. After 2 h on orbital shaking, worms were transferred onto fresh NGM plates seeded with OP50 E. coli and the pharyngeal activity was determined 2 h and 24 h later, as described above.

For the experiments with the heat-denatured variants, WT, A551P and M517R proteins at 1.5 µg/ml in 20 mM Hepes solution containing 100 mM NaCl and 1 mM $CaCl_2$ were administered to *C. elegans* (100 worms/ 100 µL) before and after incubation at 100 °C for 10 minutes. Control worms (100 worms/ 100 µL) were treated with 20 mM Hepes solution containing 100 mM NaCl and 1 mM $CaCl_2$ alone. The pharyngeal pumping was determined 2 h and 24 h after the treatment as described above.

**MD Simulations**

Starting configurations for molecular dynamics simulations were built on the basis of the crystallographic structure of full length WT GSN (PDB ID: 3FFN (3)), retaining residues M412 to L630 of chain A, comprising G4 and G5 domains. The truncation sites were capped with neutral acetylated and N-methylated termini. The structures of the three mutants A551P, E553K and M517R were generated *in silico* through the substitution of the mutated residue. Each mutant and the WT model were parameterized with the AMBER ff14SB force field; protonation states for pH 7.4 were assigned with the ProteinPrepare algorithm (63). The resulting system was solvated with TIP3P water in a 79 x 79 x 79 $Å^3$ box, neutralized and ionized with 150 mM NaCl. The preparation and building steps were conducted using the HTMD package (64).

Each system was energy-minimised for 500 steps with the L-BFGS algorithm, then equilibrated at constant pressure (NPT) at 1 atm for a total of 20 ns with the Berendsen thermostat. Given the relatively high stability of all of the variants, the dynamics at varying temperatures was probed with a set of exploratory runs of 1 μs each at 300 K, 325 K and 343 K (Figure S6), held with the Langevin thermostat.

Consistent with the thermal stability measurements, only at 343 K the simulations displayed an appreciable modulation of the local dynamics on the sampled time-scale of 1 μs. In order to gather sufficient sampling, production MD runs were carried out in triplicate for each variant and WT form at 343 K in unbiased constant-volume conditions. The 12 trajectories (4 systems x 3 replicas) were then analysed with custom Python scripts to compute donor-acceptor distances and local mean squared displacements. All of the runs employed a 4 fs time step with the hydrogen mass repartitioning scheme (65) and rigid bonds, and were computed on GPU clusters with the OpenMM 7.4.1 library (66) with the MiniOMM wrapper.

**ACKNOWLEDGEMENTS AND FUNDING SOURCES**

We acknowledge CINECA awards under the ISCRA initiative for the availability of high performance computing resources and support. T..G. thanks the participants to the GPUGRID.net project for donating computing time. The authors would like to thank Diamond Light Source for beamtime (proposal MX20221), and the staff of beamline I04 for assistance. The work was partly supported by a grant from the amyloidosis foundation to M.d.R.

**REFERENCES**

1. H. L. Yin, T. P. Stossel, Control of cytoplasmic actin gel–sol transformation by gelsolin, a calcium-dependent regulatory protein. *Nature* 281, 583–586 (1979).

2. H. Q. Sun, M. Yamamoto, M. Mejillano, H. L. Yin, Gelsolin, a multifunctional actin regulatory protein. *J. Biol. Chem.* 274, 33179–33182 (1999).

3. S. Nag, *et al.*, Ca2+ binding by domain 2 plays a critical role in the activation and stabilization of gelsolin. *Proc. Natl. Acad. Sci. U. S. A.* 106, 13713–13718 (2009).

4. H. Choe, *et al.*, The Calcium Activation of Gelsolin: Insights from the 3Å Structure of the G4–G6/Actin Complex. *Journal of Molecular Biology* 324, 691–702 (2002).

5. P. J. McLaughlin, J. T. Gooch, H. G. Mannherz, A. G. Weeds, Structure of gelsolin segment 1-actin complex and the mechanism of filament severing. *Nature* 364, 685–

692 (1993).

6. S. L. Kazmirski, *et al.*, Loss of a metal-binding site in gelsolin leads to familial amyloidosis-Finnish type. *Nat. Struct. Biol.* 9, 112–116 (2002).

7. K. Narayan, *et al.*, Activation in isolation: exposure of the actin-binding site in the C-terminal half of gelsolin does not require actin. *FEBS Lett.* 552, 82–85 (2003).

8. J. Meretoja, Familial systemic paramyloidosis with lattice dystrophy of the cornea, progressive cranial neuropathy, skin changes and various internal symptoms. A previously unrecognized heritable syndrome. *Ann. Clin. Res.* 1, 314–324 (1969).

9. A. de la Chapelle, J. Kere, G. H. Sack Jr, R. Tolvanen, C. P. Maury, Familial amyloidosis, Finnish type: G654----a mutation of the gelsolin gene in Finnish families and an unrelated American family. *Genomics* 13, 898–901 (1992).

10. C. P. Maury, E. L. Nurmiaho-Lassila, H. Rossi, Amyloid fibril formation in gelsolin-derived amyloidosis. Definition of the amyloidogenic region and evidence of accelerated amyloid formation of mutant Asn-187 and Tyr-187 gelsolin peptides. *Lab. Invest.* 70, 558–564 (1994).

11. T. Mustonen, *et al.*, Cardiac manifestations in Finnish gelsolin amyloidosis patients. *Amyloid*, 1–5 (2021).

12. S. Koskelainen, F. Zhao, H. Kalimo, M. Baumann, S. Kiuru-Enari, Severe elastolysis in hereditary gelsolin (AGel) amyloidosis. *Amyloid* 27, 81–88 (2020).

13. E.-K. Schmidt, S. Kiuru-Enari, S. Atula, M. Tanskanen, Amyloid in parenchymal organs in gelsolin (AGel) amyloidosis. *Amyloid* 26, 118–124 (2019).

14. T. Pihlamaa, T. Salmi, S. Suominen, S. Kiuru-Enari, Progressive cranial nerve involvement and grading of facial paralysis in gelsolin amyloidosis. *Muscle Nerve* 53, 762–769 (2016).

15. S. Kiuru-Enari, J. Keski-Oja, M. Haltia, Cutis laxa in hereditary gelsolin amyloidosis. *British Journal of Dermatology* 152, 250–257 (2005).

16. S. Kiuru-Enari, H. Somer, A.-M. Seppäläinen, I.-L. Notkola, M. Haltia, Neuromuscular Pathology in Hereditary Gelsolin Amyloidosis. *Journal of Neuropathology & Experimental Neurology* 61, 565–571 (2002).

17. T. Pihlamaa, S. Suominen, S. Kiuru-Enari, M. Tanskanen, Increasing amount of amyloid are associated with the severity of clinical features in hereditary gelsolin (AGel) amyloidosis. *Amyloid* 23, 225–233 (2016).

18. J. P. Solomon, L. J. Page, W. E. Balch, J. W. Kelly, Gelsolin amyloidosis: genetics, biochemistry, pathology and possible strategies for therapeutic intervention. *Crit. Rev. Biochem. Mol. Biol.* 47, 282–296 (2012).

19. S. Sethi, *et al.*, Clinical, biopsy, and mass spectrometry findings of renal gelsolin amyloidosis. *Kidney Int.* 91, 964–971 (2017).

20. S. Sethi, *et al.*, Renal amyloidosis associated with a novel sequence variant of gelsolin. *Am. J. Kidney Dis.* 61, 161–166 (2013).

21. Y. A. Efebera, *et al.*, Novel gelsolin variant as the cause of nephrotic syndrome and renal amyloidosis in a large kindred. *Amyloid* 21, 110–112 (2014).

22. G. Ratnaswamy, M. E. Huff, A. I. Su, S. Rion, J. W. Kelly, Destabilization of Ca2+-free gelsolin may not be responsible for proteolysis in Familial Amyloidosis of Finnish Type. *Proc. Natl. Acad. Sci. U. S. A.* 98, 2334–2339 (2001).

23. R. L. Isaacson, A. G. Weeds, A. R. Fersht, Equilibria and kinetics of folding of gelsolin domain 2 and mutants involved in familial amyloidosis-Finnish type. *Proc. Natl. Acad. Sci. U. S. A.* 96, 11247–11252 (1999).

24. T. Giorgino, *et al.*, Nanobody interaction unveils structure, dynamics and proteotoxicity of the Finnish-type amyloidogenic gelsolin variant. *Biochim. Biophys. Acta Mol. Basis Dis.* 1865, 648–660 (2019).

25. F. Bonì, *et al.*, Molecular basis of a novel renal amyloidosis due to N184K gelsolin variant. *Sci. Rep.* 6, 33463 (2016).

26. M. de Rosa, *et al.*, The structure of N184K amyloidogenic variant of gelsolin highlights the role of the H-bond network for protein stability and aggregation properties. *Eur. Biophys. J.* 49, 11–19 (2020).

27. C. D. Chen, *et al.*, Furin initiates gelsolin familial amyloidosis in the Golgi through a defect in Ca(2+) stabilization. *EMBO J.* 20, 6277–6287 (2001).

28. J. P. Solomon, *et al.*, The 8 and 5 kDa fragments of plasma gelsolin form amyloid fibrils by a nucleated polymerization mechanism, while the 68 kDa fragment is not amyloidogenic. *Biochemistry* 48, 11370–11380 (2009).

29. F. Bonì, *et al.*, Gelsolin pathogenic Gly167Arg mutation promotes domain-swap dimerization of the protein. *Hum. Mol. Genet.* 27, 53–65 (2018).

30. H. Zorgati, *et al.*, The role of gelsolin domain 3 in familial amyloidosis (Finnish type). *Proc. Natl. Acad. Sci. U. S. A.* 116, 13958–13963 (2019).

31. M. Sridharan, *et al.*, A Patient With Hereditary ATTR and a Novel AGel p.Ala578Pro Amyloidosis. *Mayo Clin. Proc.* 93, 1678–1682 (2018).

32. J. Cabral-Macias, *et al.*, Clinical, histopathological, and in silico pathogenicity analyses in a pedigree with familial amyloidosis of the Finnish type (Meretoja syndrome) caused by a novel gelsolin mutation. *Mol. Vis.* 26, 345–354 (2020).

33. M. Potrč, *et al.*, Clinical and Histopathological Features of Gelsolin Amyloidosis Associated with a Novel Variant p.Glu580Lys. *Int. J. Mol. Sci.* 22 (2021).

34. K. Z. Oregel, *et al.*, Atypical Presentation of Gelsolin Amyloidosis in a Man of African Descent with a Novel Mutation in the Gelsolin Gene. *Am. J. Case Rep.* 19, 374–381 (2018).

35. X. Feng, H. Zhu, T. Zhao, Y. Hou, J. Liu, A new heterozygous G duplicate in exon1 (c.100dupG) of gelsolin gene causes Finnish gelsolin amyloidosis in a Chinese family. *Brain Behav.* 8, e01151 (2018).

36. Y. Jiang, *et al.*, Analyses Mutations in GSN, CST3, TTR, and ITM2B Genes in Chinese Patients With Alzheimer's Disease. *Frontiers in Aging Neuroscience* 12 (2020).

37. C. M. Ida, *et al.*, Pituicytoma with gelsolin amyloid deposition. *Endocr. Pathol.* 24, 149–155 (2013).

38. L. Obici, G. Merlini, Seek and You Shall Find: Is Subclinical Amyloid More Common Than Expected? *Mayo Clin. Proc.* 93, 1546–1548 (2018).

39. J. Song, *et al.*, PROSPER: an integrated feature-based tool for predicting protease substrate cleavage sites. *PLoS One* 7, e50300 (2012).

40. J. Clarke, A. R. Fersht, Engineered disulfide bonds as probes of the folding pathway of barnase: increasing the stability of proteins against the rate of denaturation. *Biochemistry* 32, 4322–4329 (1993).

41. A. C. Tsolis, N. C. Papandreou, V. A. Iconomidou, S. J. Hamodrakas, A Consensus Method for the Prediction of "Aggregation-Prone" Peptides in Globular Proteins. *PLoS ONE* 8, e54175 (2013).

42. M. Stravalaci, *et al.*, The Anti-Prion Antibody 15B3 Detects Toxic Amyloid-β Oligomers. *J. Alzheimers. Dis.* 53, 1485–1497 (2016).

43. Y. Zeinolabediny, *et al.*, HIV-1 matrix protein p17 misfolding forms toxic amyloidogenic assemblies that induce neurocognitive disorders. *Sci. Rep.* 7, 10313 (2017).

44. L. Diomede, *et al.*, Cardiac Light Chain Amyloidosis: The Role of Metal Ions in Oxidative Stress and Mitochondrial Damage. *Antioxid. Redox Signal.* 27, 567–582 (2017).

45. L. Diomede, *et al.*, A Caenorhabditis elegans-based assay recognizes immunoglobulin light chains causing heart amyloidosis. *Blood* 123, 3543–3552 (2014).

46. J. S. Richardson, D. C. Richardson, Natural beta-sheet proteins use negative design to avoid edge-to-edge aggregation. *Proc. Natl. Acad. Sci. U. S. A.* 99, 2754–2759 (2002).

47. K. S. C. Reid, P. F. Lindley, J. M. Thornton, Sulphur-aromatic interactions in proteins. *FEBS Letters* 190, 209–213 (1985).

48. M. Bollati, *et al.*, High-resolution crystal structure of gelsolin domain 2 in complex with the physiological calcium ion. *Biochem. Biophys. Res. Commun.* 518, 94–99 (2019).

49. S. Mullany, *et al.*, A novel GSN variant outside the G2 calcium-binding domain associated with Amyloidosis of the Finnish type. *Hum. Mutat.* (2021) https:/doi.org/10.1002/humu.24214.

50. T. Vitali, E. Maffioli, G. Tedeschi, M. A. Vanoni, Properties and catalytic activities of MICAL1, the flavoenzyme involved in cytoskeleton dynamics, and modulation by its CH, LIM and C-terminal domains. *Arch. Biochem. Biophys.* 593, 24–37 (2016).

51. P. Cioni, E. Gabellieri, S. Marchal, R. Lange, Temperature and pressure effects on C112S azurin: volume, expansivity, and flexibility changes. *Proteins* 82, 1787–1798 (2014).

52. V. Kumar, T. K. Chaudhuri, Spontaneous refolding of the large multidomain protein malate synthase G proceeds through misfolding traps. *J. Biol. Chem.* 293, 13270–13283 (2018).

53. G. Ramsay, E. Freire, Linked thermal and solute perturbation analysis of cooperative domain interactions in proteins. Structural stability of diphtheria toxin. *Biochemistry* 29, 8677–8683 (1990).

54. P. Cioni, Role of protein cavities on unfolding volume change and on internal dynamics under pressure. *Biophys. J.* 91, 3390–3396 (2006).

55. D. Tognotti, E. Gabellieri, E. Morelli, P. Cioni, Temperature and pressure dependence of azurin stability as monitored by tryptophan fluorescence and phosphorescence. The case of F29A mutant. *Biophys. Chem.* 182, 44–50 (2013).

56. W. Kabsch, XDS. *Acta Crystallogr. D Biol. Crystallogr.* 66, 125–132 (2010).

57. P. R. Evans, G. N. Murshudov, How good are my data and what is the resolution? *Acta Crystallographica Section D Biological Crystallography* 69, 1204–1214 (2013).

58. A. J. McCoy, *et al.*, Phaser crystallographic software. *J. Appl. Crystallogr.* 40, 658–674 (2007).

59. P. V. Afonine, *et al.*, Real-space refinement in PHENIX for cryo-EM and crystallography. *Acta Crystallogr D Struct Biol* 74, 531–544 (2018).

60. P. Emsley, B. Lohkamp, W. G. Scott, K. Cowtan, Features and development of Coot. *Acta Crystallogr. D Biol. Crystallogr.* 66, 486–501 (2010).

61. M. de Rosa, *et al.*, Decoding the Structural Bases of D76N ß2-Microglobulin High Amyloidogenicity through Crystallography and Asn-Scan Mutagenesis. *PLoS One* 10, e0144061 (2015).

62. H. LeVine 3rd, Thioflavine T interaction with synthetic Alzheimer's disease beta-amyloid peptides: detection of amyloid aggregation in solution. *Protein Sci.* 2, 404–410 (1993).

63. G. Martínez-Rosell, T. Giorgino, G. De Fabritiis, PlayMolecule ProteinPrepare: A Web Application for Protein Preparation for Molecular Dynamics Simulations. *J. Chem. Inf. Model.* 57, 1511–1516 (2017).

64. S. Doerr, M. J. Harvey, F. Noé, G. De Fabritiis, HTMD: High-Throughput Molecular Dynamics for Molecular Discovery. *J. Chem. Theory Comput.* 12, 1845–1852 (2016).

65. C. W. Hopkins, S. Le Grand, R. C. Walker, A. E. Roitberg, Long-Time-Step Molecular Dynamics through Hydrogen Mass Repartitioning. *J. Chem. Theory Comput.* 11, 1864–1874 (2015).

66. P. Eastman, *et al.*, OpenMM 7: Rapid development of high performance algorithms for molecular dynamics. *PLoS Comput. Biol.* 13, e1005659 (2017).

Supporting information

# A novel hotspot of gelsolin instability and aggregation propensity triggers a new mechanism of amyloidosis

Bollati et al.

**Supplementary figures and tables:**

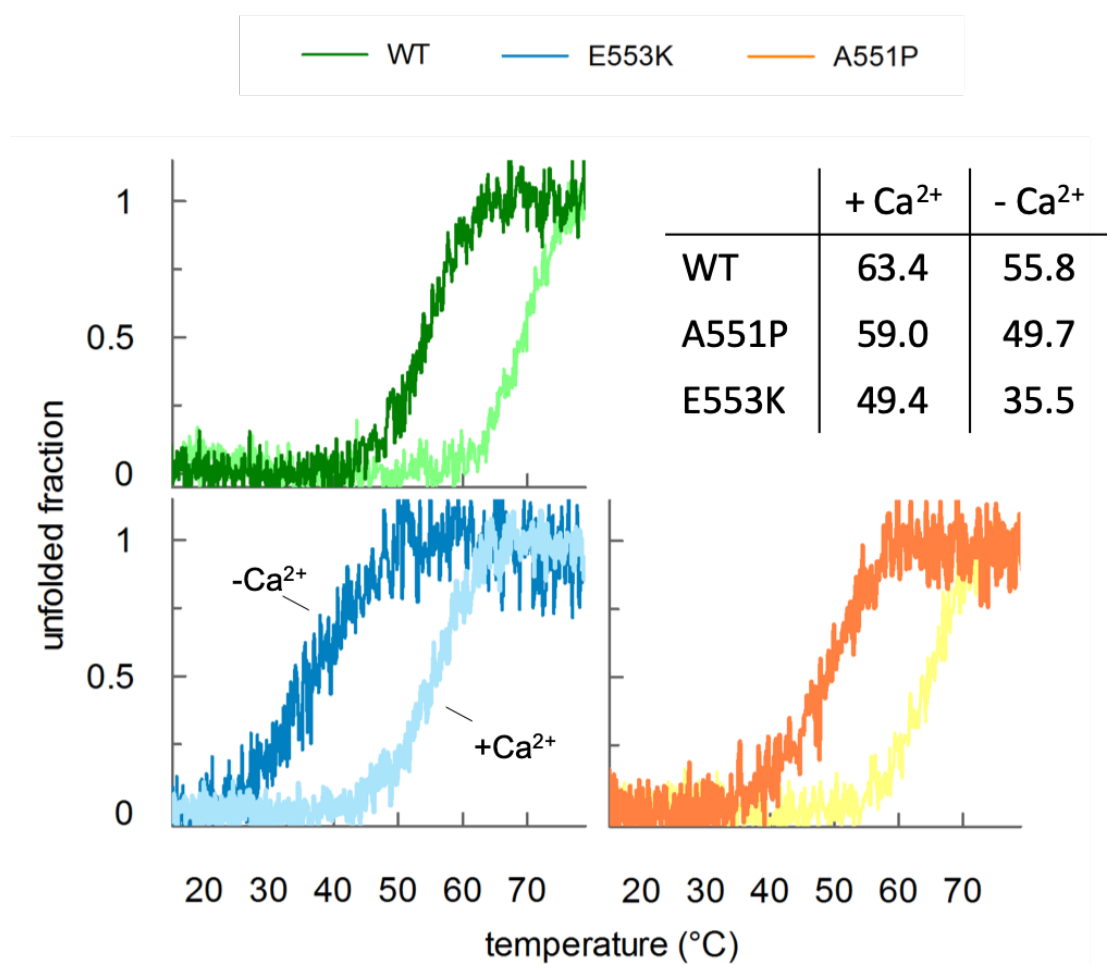

**Figure S1 Thermal stability of the isolated WT and mutated G5 domains.** Thermal denaturation was followed by circular dichroism in the far UV region in the presence (darker shade trace) and absence (light shade) of $Ca^{2+}$. The table reports the Tm values (± 0.3 °C) calculated for G5 of WT, A551P and E553K as the minima of the first derivative of the traces.

| Sequences | Domain |
|---|---|
| 42-QIWRV-46 | G1 |
| 66-DAYVILKTV-74 | G1 |
| 100-AAAIFTVQL-108 | G1 |
| 128-ATFLGY-133 | G1 |
| 157-VVVQRLF-163 | G2 |
| 186-**GDCFIL**-190 | G2 – known amyloidogenic sequence |
| 303-DCFIL-307 | G3 |
| 379-GLSYLSS-385 | G3 |
| 445-DSYIILYNY-453 | G4 |
| 461-QIIYNW-466 | G4 |
| 480-AILTA-484 | G4 |
| **508-LMSLFG-513** | G4 – helix close to M517 |
| **538-LFQV-540** | G5 – strand close to A551 and E553 strand |
| **574-AAYLWVGT-581** | G5 – strand of the same sheet |
| **594-LLRV-597** | G5 – helix close to interface |
| 650-GRFVIE-655 | G5 |
| 675-DTWDQVFVWV-684 | G6 |
| 730-VGWFL-734 | G6 |

**Table S1 Aggregation prone sequences identified with the program Amylpred2** (1). We show the hits obtained by applying the most stringent consensus (5 out of 10 predictions) that still identifies the experimentally validated sequence in G2 (SFNNGDCFILD).

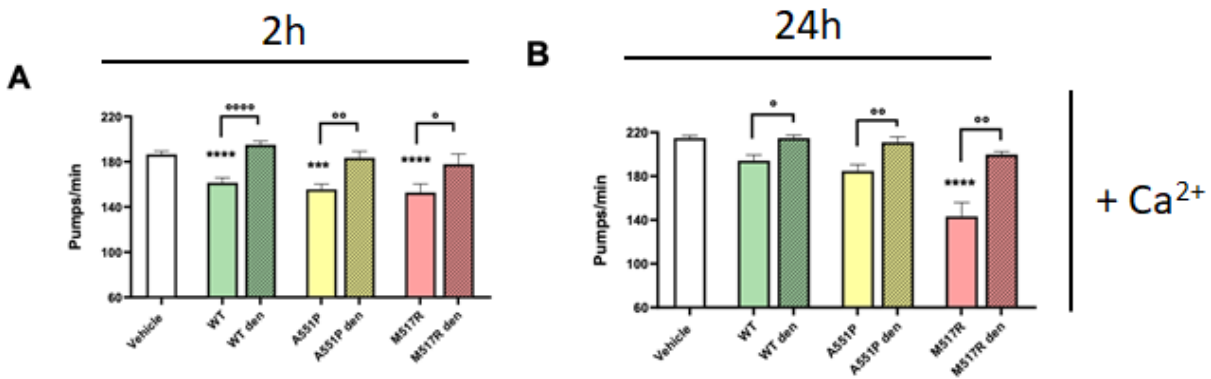

**Figure S2: Heat denatured full length GSN variants are not toxic to worms.** Proteins at 17.6 µM in 20 mM Hepes solution containing 100 mM NaCl and 1 mM $CaCl_2$ were administered to *C. elegans* (100 worms/ 100 µL) before (WT, A551P and M517R) and after incubation at 100 °C for 10 minutes (WT den, A551P den and M517R den). Control worms (100 worms/ 100 µL) were treated with 20 mM Hepes solution containing 100 mM NaCl and 1 mM $CaCl_2$ alone (Vehicle). The pharyngeal activity was determined (A) 2 h and (B) 24 h after the treatment by determining the number of pharyngeal bulb contraction (pumps/min). Data are mean ± SE (N= 20/group). ***$p<0.001$ and ****$p<0.0001$ vs Vehicle according to one-way ANOVA and Bonferroni post hoc test. °$p<0.05$, °° $p<0.01$ and °°°°$p<0.0001$ according to Student's t-test.

**Table S2 Data collection and refinement statistics.** Model and structure factors were deposited in the protein data bank under accession code 7P2B. Values in parentheses refer to the highest resolution shells.

| Data collection | |
|---|---|
| Space group and cell dimensions<br>a, b, c;<br>α, β, γ (Å;°) | P 4 $2_1$ 2<br>170, 170, 152;<br>90, 90, 90 |
| Unique reflections | 44997 (3291) |
| Resolution range (Å) | 19.91-3.0 (3.08-3.00) |
| I/σ(I) | 11.53 (1.10) |
| CC 1/2 | 1.00 (0.58) |
| Completeness (%) | 100 (99.5) |
| Multiplicity | 27.3 (28.4) |
| **Refinement** | |

| Rwork/Rfree* | 0.213/0.262 |
|---|---|
| RMSD Bonds/angles (Å/°) | 0.003/0.588 |
| Ramachandran outliers (%) | 0.57 |
| B factors (Å$^2$) § | 83.6 |
| PDB id | 7P2B |

*Rwork =$\Sigma_{hkl}$|Fo|-|Fc|/$\Sigma_{hkl}$|Fo| for all data, except 5%, which were used for calculation of the Rfree.

§Average temperature factors for the overall structure.

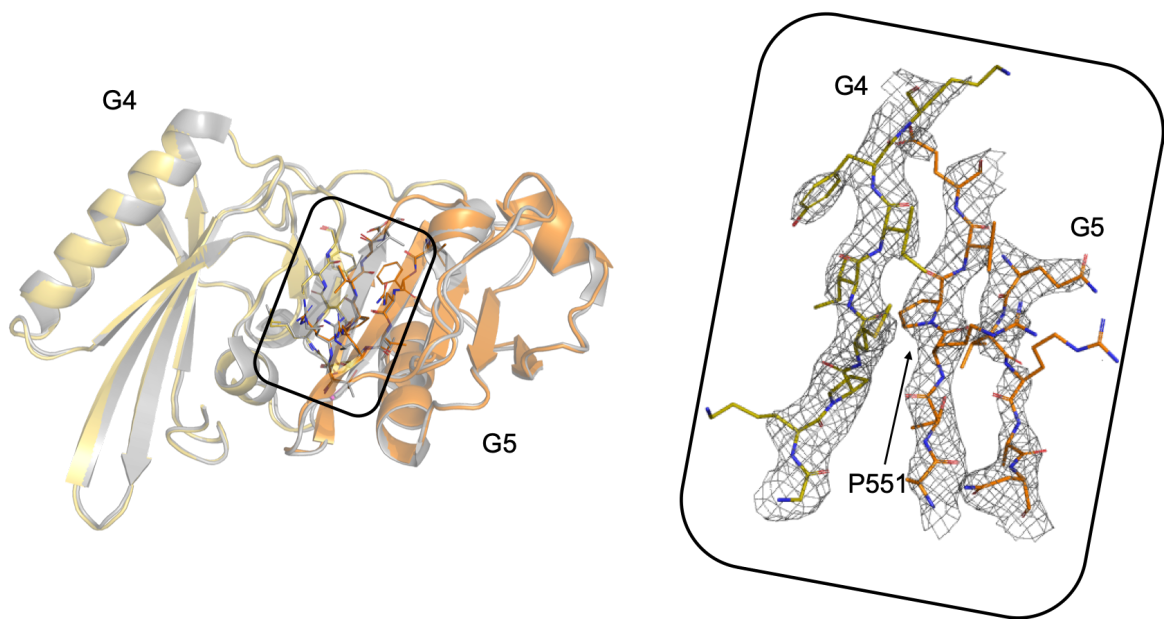

**Figure S3 Crystal structure of the A551P variant.** Superimposition of the G4G5 portion of the structure of full length GSN A551P (yellow/orange, this work) with GSN WT (grey, 3FFN (2)). Point of view, orientation and colors as in Figure 6. In the close view of the interface area, electron density (at 1.6 σ) is displayed as a grey mesh for the strand hosting the mutation and those flanking.

**Table S3 Impact of the mutation on the H-bond network tethering G4 and G5.** We computed the occupancy of each potential H-bond involved in the formation of the parallel β interface, defined as the fraction of time (0 to 1) that the corresponding donor-acceptor atom distance was <3 Å over the 1 µs simulations; values are averages and standard deviations of 3 runs of 1 µs. ΔWT indicates the difference with respect to WT along with its error computed by error propagation. For a graphical map of the interactions object of this analysis see Figure S4. In A551P we observed the loss of bond between atoms O-517 and N-551, due to the absence of the NH group in the proline, and a weakening of the O-551 and N-519 bond, with a slight 13% occupancy reduction. As hypothesized, even the adjacent residue in position 550 is located at a distance greater than 3 Å from R542 (O-550/N-542) for 39% of simulated time. The E553K substitution causes the loss of the binding between the carboxyl group of E and the amide group of K521 and the consequent destabilization of the H-bond with I519 (27% occupancy reduction). Conversely, M517R mutation demonstrates a lower incidence in the loss of hydrogen bonds between the two β-strands. Only a loss of 14% of the binding occupancy at the interface, involving O-517 and N-551, as well as a loss of 13% between interface atoms O-551 and N-519 were detected. Interestingly, residue I519 shows a significant reduction in occupancy in A551P and M517R mutants unlike in the WT, where it is very conserved. This may suggest a key role in keeping the interface between the two domains stable. Legend: O and N, backbone oxygen and nitrogen atoms; OE, sidechain oxygen at position ε; NZ, sidechain nitrogen at position ζ.

| | | WT | | | A551P | | | | | | E553K | | | | | | M517R | | | | | |
|---|---|---|---|---|---|---|---|---|---|---|---|---|---|---|---|---|---|---|---|---|---|---|
| | | H-bond occupancy | | | H-bond occupancy | | | ΔWT | | | H-bond occupancy | | | ΔWT | | | H-bond occupancy | | | ΔWT | | |
| O_439 | N_518 | 0.43 | ± | 0.06 | 0.45 | ± | 0.06 | 0.01 | ± | 0.09 | 0.37 | ± | 0.07 | -0.07 | ± | 0.09 | 0.41 | ± | 0.11 | -0.02 | ± | 0.12 |
| O_517 | N_551 | 0.86 | ± | 0.02 | No NH- group | | | **-0.86** | ± | **0.02** | 0.78 | ± | 0.02 | -0.08 | ± | 0.03 | **0.71** | ± | **0.15** | **-0.14** | ± | **0.15** |
| O_549 | N_517 | 0.38 | ± | 0.08 | 0.48 | ± | 0.02 | 0.09 | ± | 0.08 | 0.32 | ± | 0.06 | -0.06 | ± | 0.10 | 0.43 | ± | 0.12 | 0.05 | ± | 0.15 |
| O_551 | N_519 | 0.70 | ± | 0.01 | **0.57** | ± | **0.07** | **-0.13** | ± | **0.07** | 0.71 | ± | 0.04 | 0.01 | ± | 0.04 | **0.58** | ± | **0.13** | **-0.13** | ± | **0.13** |
| O_519 | N_553 | 0.49 | ± | 0.03 | 0.45 | ± | 0.05 | -0.04 | ± | 0.06 | **0.23** | ± | **0.02** | **-0.27** | ± | **0.04** | 0.49 | ± | 0.09 | 0.00 | ± | 0.10 |
| OE_553 | N_521 | 0.66 | ± | 0.06 | 0.60 | ± | 0.04 | -0.06 | ± | 0.07 | No COO- group | | | **-0.66** | ± | **0.06** | 0.68 | ± | 0.12 | 0.02 | ± | 0.14 |
| OE_553 | NZ_521 | 0.04 | ± | 0.01 | 0.04 | ± | 0.02 | 0.00 | ± | 0.02 | No COO- group | | | -0.04 | ± | 0.01 | 0.05 | ± | 0.06 | 0.02 | ± | 0.06 |
| O_538 | N_554 | 0.62 | ± | 0.01 | 0.61 | ± | 0.03 | -0.01 | ± | 0.03 | 0.54 | ± | 0.04 | -0.09 | ± | 0.04 | 0.62 | ± | 0.05 | 0.00 | ± | 0.05 |
| O_552 | N_540 | 0.77 | ± | 0.02 | 0.77 | ± | 0.02 | -0.01 | ± | 0.03 | 0.76 | ± | 0.01 | -0.02 | ± | 0.02 | 0.72 | ± | 0.05 | -0.05 | ± | 0.05 |
| O_540 | N_552 | 0.57 | ± | 0.00 | **0.71** | ± | **0.02** | **0.14** | ± | **0.02** | 0.57 | ± | 0.03 | 0.00 | ± | 0.03 | 0.51 | ± | 0.05 | -0.06 | ± | 0.05 |
| O_542 | N_550 | 0.66 | ± | 0.02 | 0.67 | ± | 0.05 | 0.01 | ± | 0.05 | 0.67 | ± | 0.03 | 0.02 | ± | 0.04 | 0.62 | ± | 0.03 | -0.03 | ± | 0.04 |
| O_550 | N_542 | 0.49 | ± | 0.08 | **0.10** | ± | **0.03** | **-0.39** | ± | **0.08** | 0.56 | ± | 0.02 | 0.07 | ± | 0.08 | **0.62** | ± | **0.13** | **0.14** | ± | **0.15** |

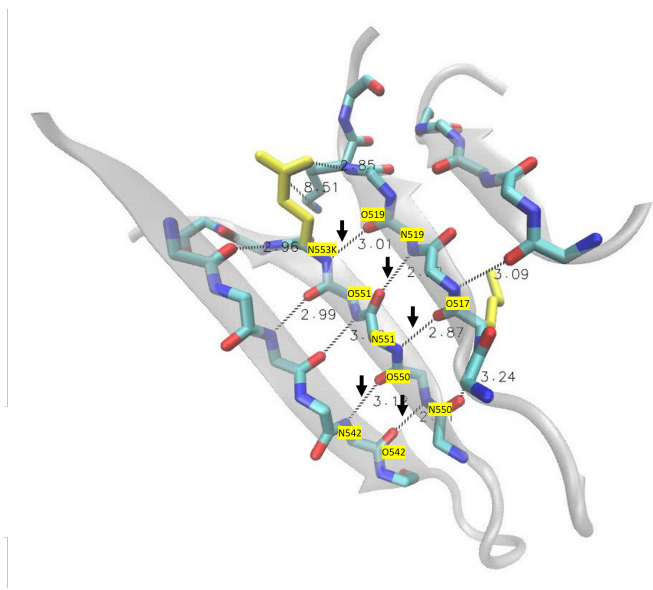

**Figure S4: H-bond network at the G4:G5 interface.** G4 and G5 domains form a continuous β-sheet with a large number of intimate contacts. Represented here are the H-bonds whose quantification of their occupancy along the molecular dynamics simulations is reported in Table S3.

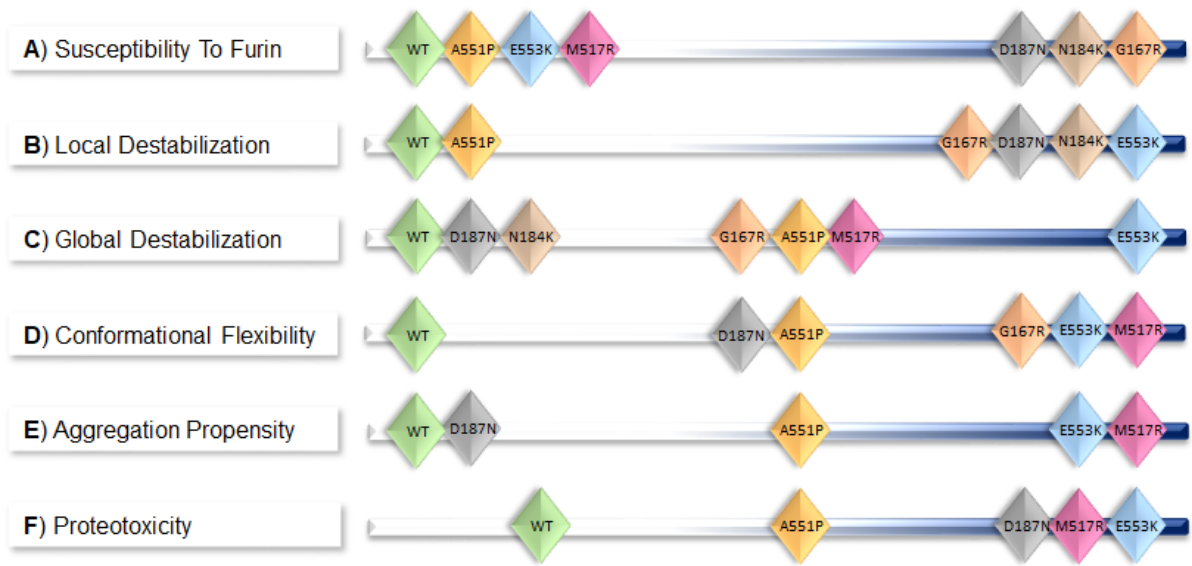

*Figure S5: Ranking of the molecular features of so far characterized GSN amyloidogenic variants*. G2 and G4:G5 variants are semi-quantitatively scored with respect to the WT protein, reviewing data from literature and this work. (**A**) Furin proteolysis assays show that only the G2-linked variants (D187N/Y, N184K and G167R) cause the exposure of a furin-sensitive site and trigger the canonical aggregation mechanism (3–5). (**B-C**) Local destabilization is that measured on the isolated G2 or G5 domains; whereby global, when it is measurable on the FL proteins. G2-linked mutations, with the exception of G167R, are mainly responsible for a local destabilization (4–8), while all G4:G5 mutations cause a global destabilization effect, suggesting an impairment of the inter-domains connectivity. In the E553K, M517R and G167R variants we observe a combination of local destabilization and global rearrangements. (**D**) Conformational flexibility is inferred by MD experiments on the G2G3 portion of the protein for D187N/Y and G167R (3) and on the G4G5 stretch for the others. (**E**) ThT assays show that D187N, as representative of the G2-linked variants, does not aggregate in its uncleaved FL form (9), while G4:G5 variants show some propensity to form ThT-sensitive species. (**F**) This dataset is scaled on the vehicle rather than the WT protein to underline its toxicity measurable in the *C. elegans* assay. However, all pathological variants are significantly more toxic than the WT. The E553K variant shows two different behaviors: a mild transitory proteotoxicity which turns into the highest permanent effect of all the studied variants upon aging (Figure 4).

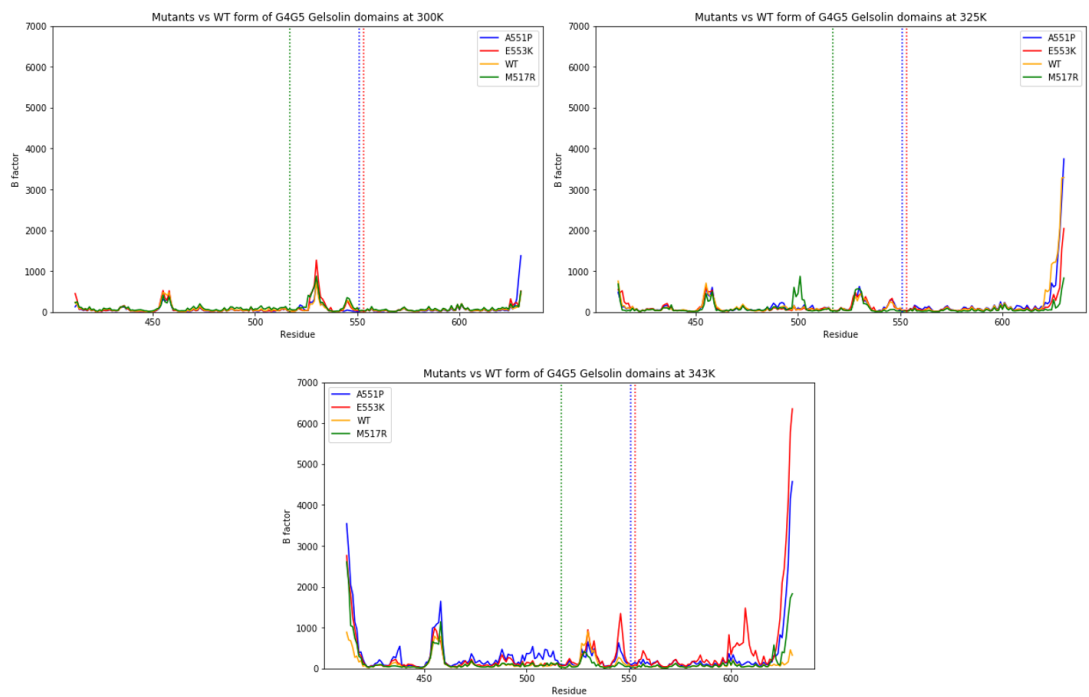

**Figure S6 Temperature-dependent conformational flexibility of mutated GSN.** Displacements of atoms of the WT and G4:G5 mutated GSN were evaluated along the trajectories of molecular dynamics simulations and expressed as B factors. Values at the following temperature were computed independently 300, 325 and 343 K.

## Supplementary references

1. A. C. Tsolis, N. C. Papandreou, V. A. Iconomidou, S. J. Hamodrakas, A Consensus Method for the Prediction of "Aggregation-Prone" Peptides in Globular Proteins. *PLoS ONE* **8**, e54175 (2013).

2. S. Nag, *et al.*, Ca2+ binding by domain 2 plays a critical role in the activation and stabilization of gelsolin. *Proc. Natl. Acad. Sci. U. S. A.* **106**, 13713–13718 (2009).

3. H. Zorgati, *et al.*, The role of gelsolin domain 3 in familial amyloidosis (Finnish type). *Proc. Natl. Acad. Sci. U. S. A.* **116**, 13958–13963 (2019).

4. M. E. Huff, L. J. Page, W. E. Balch, J. W. Kelly, Gelsolin domain 2 Ca2+ affinity determines susceptibility to furin proteolysis and familial amyloidosis of finnish type. *J. Mol. Biol.* **334**, 119–127 (2003).

5. F. Bonì, *et al.*, Molecular basis of a novel renal amyloidosis due to N184K gelsolin variant. *Sci. Rep.* **6**, 33463 (2016).

6. S. L. Kazmirski, *et al.*, Loss of a metal-binding site in gelsolin leads to familial amyloidosis-Finnish type. *Nat. Struct. Biol.* **9**, 112–116 (2002).

7. R. L. Isaacson, A. G. Weeds, A. R. Fersht, Equilibria and kinetics of folding of gelsolin domain 2 and mutants involved in familial amyloidosis-Finnish type. *Proc. Natl. Acad. Sci. U. S. A.* **96**, 11247–11252 (1999).

8. F. Bonì, *et al.*, Gelsolin pathogenic Gly167Arg mutation promotes domain-swap dimerization of the protein. *Hum. Mol. Genet.* **27**, 53–65 (2018).

9. J. P. Solomon, *et al.*, The 8 and 5 kDa fragments of plasma gelsolin form amyloid fibrils by a nucleated polymerization mechanism, while the 68 kDa fragment is not amyloidogenic. *Biochemistry* **48**, 11370–11380 (2009).